\RequirePackage[2020-02-02]{latexrelease}
\documentclass[twocolumn, switch]{article} 

\usepackage{preprint}

\usepackage{amsmath, amsthm, amssymb, amsfonts}

\usepackage[numbers,square,sort&compress]{natbib}
\usepackage[utf8]{inputenc}	
\usepackage[T1]{fontenc}	
\usepackage{xcolor}		
\usepackage[colorlinks = true,
            linkcolor = purple,
            urlcolor  = blue,
            citecolor = cyan,
            anchorcolor = black]{hyperref}	
\usepackage{booktabs} 		
\usepackage{nicefrac}		
\usepackage{microtype}		
\usepackage{lineno}		
\usepackage{float}			

\usepackage{lipsum}		

\usepackage{bm}
\usepackage{graphicx}
\graphicspath{{./images/}}
\usepackage{multirow}
\usepackage{array}
\usepackage{makecell}
\usepackage{url}

\usepackage{newfloat}
\DeclareFloatingEnvironment[name={Supplementary Figure}]{suppfigure}
\usepackage{sidecap}
\sidecaptionvpos{figure}{c}

\usepackage{titlesec}
\titlespacing\section{0pt}{12pt plus 3pt minus 3pt}{1pt plus 1pt minus 1pt}
\titlespacing\subsection{0pt}{10pt plus 3pt minus 3pt}{1pt plus 1pt minus 1pt}
\titlespacing\subsubsection{0pt}{8pt plus 3pt minus 3pt}{1pt plus 1pt minus 1pt}

\usepackage{tikz,xcolor,hyperref}

\definecolor{lime}{HTML}{A6CE39}
\DeclareRobustCommand{\orcidicon}{
	\begin{tikzpicture}
	\draw[lime, fill=lime] (0,0) 
	circle [radius=0.16] 
	node[white] {{\fontfamily{qag}\selectfont \tiny ID}};
	\draw[white, fill=white] (-0.0625,0.095) 
	circle [radius=0.007];
	\end{tikzpicture}
	\hspace{-2mm}
}
\foreach \x in {A, ..., Z}{\expandafter\xdef\csname orcid\x\endcsname{\noexpand\href{https://orcid.org/\csname orcidauthor\x\endcsname}
			{\noexpand\orcidicon}}
}

\title{Eigen-spectrograms: an interpretable feature space for bearing fault diagnosis based on artificial intelligence and image processing}

\usepackage{xwatermark}
\usepackage{catoptions}
\newwatermark[allpages,color=gray!60,angle=90,scale=0.35, xpos=3.9in,ypos=0, width=1500pt]{This is an original manuscript of an article published by Taylor \& Francis in Mechanics of Advanced Materials and Structures on 22 July 2022, available online: \url{http://www.tandfonline.com/10.1080/15376494.2022.2102274}.}
\newwatermark[firstpage,color=gray!90,angle=0,scale=0.28, xpos=0in,ypos=-5in]{*correspondence: \texttt{luigi.dimaggio@polito.it}}

\usepackage{authblk}

\author[1]{Eugenio Brusa\orcidA{}}
\author[1]{Cristiana Delprete\orcidB{}}
\author[1]{Luigi Gianpio Di Maggio\orcidC{}}

\affil[1]{Department of Mechanical and Aerospace Engineering (DIMEAS) \protect \\ Politecnico di Torino, Corso Duca degli Abruzzi 24, 10129 Torino, Italy}

\begin{document}
\twocolumn[ 
  \begin{@twocolumnfalse} 
  
\maketitle

\begin{abstract}
The Intelligent Fault Diagnosis of rotating machinery currently proposes some captivating challenges. Although results achieved by artificial intelligence and deep learning constantly improve, this field is characterized by several open issues. Models’ interpretation is still buried under the foundations of data driven science, thus requiring attention to the development of new opportunities also for machine learning theories. This study proposes a machine learning diagnosis model, based on intelligent spectrogram recognition, via image processing. The approach is characterized by the employment of the eigen-spectrograms and randomized linear algebra in fault diagnosis. Randomized algebra and eigen-spectrograms enable the construction of a significant feature space, which nonetheless emerges as a viable device to explore models’ interpretations. The computational efficiency of randomized approaches provides reading keys of well-established statistical learning theories such as the Support Vector Machine (SVM). Machine learning applied to spectrogram recognition shows to be extremely accurate and efficient as compared to state of the art results.
\end{abstract}
\vspace{0.35cm}

  \end{@twocolumnfalse} 
] 



\section{Introduction}
The growing complexity of industrial rotating systems has found in predictive maintenance strategies and condition monitoring techniques some key assets to enhance the production performance, by reducing maintenance costs, machine failures and downtimes \cite{Mohanty2014Machinery}.

The problem of developing some robust monitoring techniques for condition-based maintenance has gradually flanked the crucial issue of building reliable models able to estimate wear and fatigue of such complex systems, for scheduled maintenance. In particular, Rolling Element Bearings (REB) are among the most critical components in industrial rotating machinery, since their durability suffers of a wide statistical dispersion \cite{Randall2011Vibration-based}, which is one of the prominent aspects that makes time-based maintenance approaches unadvisable. Thus, the Remaining Useful Life (RUL) assessment and the machine health management based on current condition \cite{Brusa2020Design,Brusa2021Thermal,Gougam2021Fault,Wang2018Prognostics} are clearly more reliable than scheduled maintenance models.

Some remarkable scientific efforts carried out in the last decades have brought to light many signal processing tools for REB vibration analysis, relying on physics-based speculations. For instance, it is worth mentioning the great attention that has been paid, since the 1980s, to amplitude demodulation by means of the envelope analysis \cite{McFadden1984Vibration}, whose effectiveness as a diagnostic tool was widely proven over the past decades \cite{Abboud2019Advanced,Brusa2020Health,Delprete2005Rolling,Delprete2020Bearing,Randall2011Rolling,Sawalhi2008Semi-automated,Zhong2022Fault}. Also, no less research was inspired by the consequent issues which have arisen for the choice of optimal demodulation bands in non-stationary signals \cite{Abboud2017Envelope, Daga2021Fast,Moshrefzadeh2018Autogram:,RohaniBastami2020Rolling,Smith2019Optimal,Wang2019Blind,Yang2019Rolling}.

In parallel to the physics-based paradigms, which have led science and engineering from its early days, the past decade has seen the rapid development of data driven science in many engineering fields. Experts’ knowledge of physical phenomena underpinning human-inferred models is replaced in these approaches by learning algorithms, which apprehend from training data, shaping themselves based on some received inputs. In this sense, the cost for the detachment from physical assumptions paid with a loss in models’ interpretability is balanced by the opportunity of accurately foreseeing the behavior of complex systems. The possible benefits of data driven models for discovering hidden path in vibration data have been explored by numerous investigations of which a deep insight is provided in some recent review papers of Liu et al. \cite{Liu2018Artificial}, Zhao et al. \cite{Zhao2019Deep} and Lei et al. \cite{Lei2020Applications}.

The Intelligent Fault Diagnosis (IFD) refers to the use of machine learning and deep learning algorithms, respectively, for machine health assessment \cite{Lei2020Applications}. Machine learning approaches such as Support Vector Machine (SVM) \cite{Cortes1995Support-vector, Widodo2007Support} and k-Nearest Neighbor (kNN) \cite{Cover1967Nearest} have dominated the scene of Artificial Intelligence (AI) and IFD until mid-2010s. For instance, multi-class SVM was combined with wavelet analysis \cite{Xian2009intelligent}, Empirical Mode Decomposition (EMD) \cite{Yang2007fault}, multiscale approaches \cite{Hao2011Application}, Particle Swarm Optimization (PSO) \cite{Liu2013Multi-fault,Zhong2022Fault} and envelope analysis \cite{Guo2009Rolling}. The main goal of proposed strategies, which were combined with SVM over the past years, was to provide some signal processing tools capable of extracting significant features \cite{Sharma2016Feature,Xi2000Bearing} to input in the training process. Indeed, machine learning methods need human assistance for feature extraction. This is the case of Sharma et al. \cite{Sharma2016Feature}, who employed time domain features (e.g. skewness, kurtosis, crest factor) to input an SVM classifier. Instead, EMD features were used by Yang et al. \cite{Yang2007fault} who extracted amplitude ratios at fault characteristic frequencies in the Intrinsic Mode Function (IMF) envelope spectrum.   

By converse, deep learning \cite{Li2020Raw,Lv2021Fault} and Deep Neural Networks profit from deep multi-layered and hierarchical architectures, to perform the automated features extraction and to infer outputs, from very large datasets, typical of modern IoT systems. Anyway, it is worth underlining that most of the benchmark REB datasets \cite{0CWRU,Daga2019Politecnico,Nectoux2012PRONOSTIA,Qiu2006Wavelet,Smith2015Rolling} available in literature require data augmentation in order to be treated in a IFD perspective, while remaining still far from standard image recognition datasets, including millions of samples\cite{Krizhevsky2017Imagenet}.

Even though the theoretical background of Neural Networks (NNs) traces its roots in the twentieth century \cite{Fukushima1982Neocognitron}, the milestone of Convolutional Neural Networks (CNNs) was recently defined due to computer vision tasks. An example of these is the classification of very large images datasets such as ImageNet \cite{Krizhevsky2017Imagenet} in 2012. Notably, NNs were not yet included in the top ten data mining algorithms published in 2008\cite{Wu2008Top}.

Then, in light of the striking attractiveness of embedded feature extraction, deep learning became one of the most popular methods also for IFD thanks to stacked autoencoders (AE) approaches \cite{Alabsi2021Bearing,Guo2017Deep,Liu2016Rolling} and CNN-based methodologies \cite{Li2020Raw}. An autoencoder is built by assembling an encoder network with a decoder network. The encoder maps input data in a latent feature space, whereas the encoder reconstructs input data from the latent space. Also, autoencoders can be stacked to achieve deeper and more complex representations of fault features. Generally, the training of such architectures is unsupervised and it results in an automated feature extraction process. Guo et al. \cite{Guo2017Deep} employed autoencoders for the feature extraction and the denoising of gearbox and bearings signals, whereas Alabsi et al. \cite{Alabsi2021Bearing} showed that some pre-processing is still needed to improve the performances of autoencoders. On the other hand, CNNs take advantage of supervised learning and they are able to preserve the spatial information of data by performing multidimensional filtering. The latter property is particularly suitable for images and it paves the way for employing time-frequency images as training data. Therefore, CNN-based methods either act by pre-processing data with the purpose of adapting them to CNNs, essentially designed for images \cite{Guo2018novel,Islam2019Automated,Sun2017intelligent,Wen2018New}, or directly work on greyscale and infrared images.

A remarkable research was performed in this field, using vibration data and infrared images in CNNs for rotor systems monitoring \cite{Janssens2018Deep,Yuan2018novel}, whereas images and CNNs were employed also in the health monitoring of balancing tail ropes by Zhou et al. \cite{Zhou2018Health}. Then, Yoo and Baek \cite{Yoo2018novel} obtained promising results in RUL estimation by constructing a CNNs health indicator trained with wavelet time-frequency representations. Shao et al. \cite{Shao2019Highly}, instead, developed a CNN methodology based on Transfer Learning (TL) \cite{Brusa2021Deep,Cao2018Preprocessing-Free,Xiao2019Transfer,Zhang2017Transfer,Zhang2021intelligent} which is considered one of the future perspective for research on Deep Learning IFD \cite{Lei2020Applications}.

The application of data-driven tools to machine diagnosis is often limited by the  opportunity to generalize the diagnosis model to unseen working conditions. Nonetheless, the presence of discrepancies between training and test data represents one of the major concerns for industrial applications. Recent works investigated this issue. Han et al. \cite{Han2021} proposed an hybrid generalization network by employing adversarial learning and achieved improved generalization abilities for gearbox datasets. Also, unseen working conditions were analyzed by using spatiotemporal feature learning \cite{Han2019}. Instead, Zhiyi et al. \cite{Zhiyi2020} took advantage of autoencoders to transfer diagnosis models between different machines.

\section{Intelligent Fault Diagnosis: Open issues, aim and motivation}

The automated feature extraction through deep learning involves two main issues. The first concern is related to the analysis of small datasets, provided that thousands samples may already be considered a limited datasets for neural networks. Namely, deep architectures are endowed with a very large number of parameters to be optimized with respect to machine learning algorithms. Thereby, it is urgent to feed Deep NNs with a huge volume of labeled data to prevent overfitting which would lead to models perfectly accommodating training data but still hardly-generalizable. For instance, Yang et al. \cite{Yang2007fault} and Sharma et al. \cite{Sharma2016Feature} showed that SVM can achieve better generalization capabilities with respect to neural networks when small datasets are analyzed. A large amount of labeled data is not always available for machinery diagnostics and the construction of standardized datasets is becoming a challenging research field.

TL can overcome the issues related to insufficient labeled data by transferring the knowledge acquired in certain engineering scenarios to similar ones through pretrained networks. Indeed, it was showed that wavelet time-frequency images of damaged bearings can be accurately classified by adopting CNNs pretrained on ImageNet \cite{Shao2019Highly}, and by using deep architectures pre-trained for sound event detection \cite{Brusa2021Deep}.

Secondly,  the end-to-end diagnostic capabilities offered by Deep NNs impact on the interpretability of AI models, since the links between input features and model predictions are buried under a black-box. Research on statistical learning theories such as SVM is therefore encouraged and deserve further improvements \cite{Lei2020Applications}, since features and parameters provide better interpretability.

Some early image processing tools were already proposed in 2014 \cite{Klein2014Bearing}, but they did not yet take advantage of the AI algorithms, that were gaining attention in those years. Indeed, applications of AI image recognition to signal processing tools would mature few years later \cite{Shao2017Rolling,Wen2018New,Yoo2018novel}. Although images have been widely used for feeding convolutional networks, manual feature extraction and machine learning for image-based fault diagnosis have been little investigated.

This study proposes a methodology for bearing fault diagnosis, based on spectrogram image processing. 
\begin{figure*}[h]
    \centering
    \includegraphics[scale=0.6]{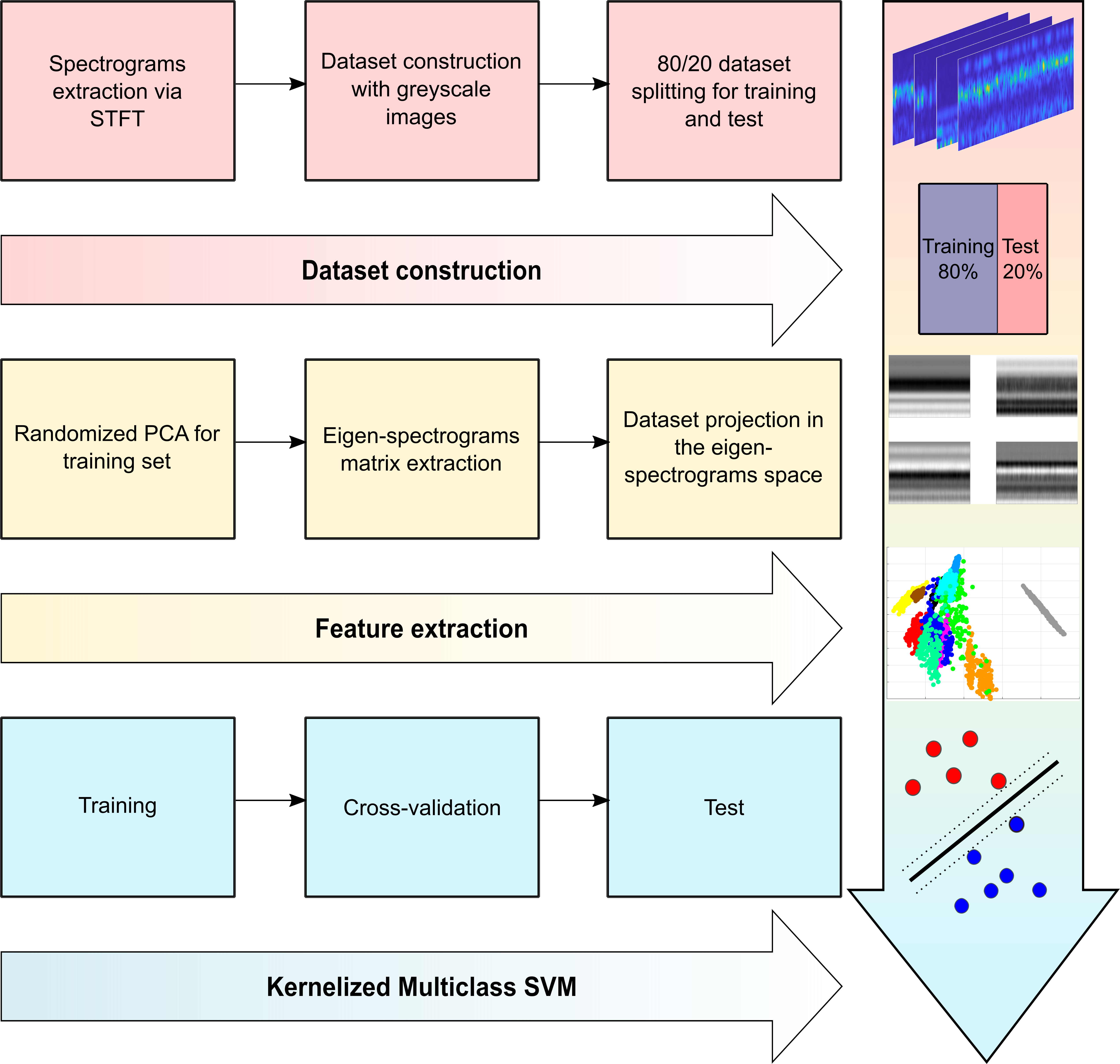}
    \caption{Spectrograms image processing via rPCA and SVM.}
    \label{fig:1}
\end{figure*}
The proposed approach (Fig.\ref{fig:1}) is able to accurately detect bearing faults and classify their type and severity by means of AI spectrogram recognition, also in presence of noisy data. Furthermore, this paper presents a reading key to link features contribution to model results, that is interpretability. One of the core parts is embodied by the implementation of the so-called \lq Randomized Linear Algebra\rq (RLA) for intelligent recognition of signal processing outcomes such as spectrograms. RLA is acknowledged as one of the cornerstones of modern data science since it represents an extremely streamlined data mining tool for extracting dominant low-rank structures underlying big datasets \cite{Brunton2019Data}. Given that these are expected to populate machinery IFD in the upcoming IoT era \cite{Lei2020Applications,Liu2018Artificial,Zhao2019Deep}, Randomized Algebra may likewise arise as a groundbreaking engineering tool.

Data matrices are projected in a principal components (PCs) subspace where most of the data variance is enclosed. This low-dimensional representation of the dataset is undertaken as feature space for a multiclass SVM classifier with polynomial kernel. Notably, the above-mentioned subspace is identified by principal components with a concise interpretation since they are images and, above all, spectrograms. The authors of this work propose the term \emph{eigen-spectrograms} for referring to those. The terminology is inspired by facial recognition science, where similar research was performed with the definition of the word \emph{eigenfaces} \cite{Belhumeur1997Eigenfaces,Brunton2019Data,Kirby1990Application,Sirovich1987Low-dimensional,Turk1991Eigenfaces}, denoting the dominant correlations between face images.

Namely, eigen-spectrograms provide the opportunity to visualize the knowledge acquired through the Randomized Principal Component Analysis (rPCA). Such a peculiarity is strictly connected to the application of a machine learning tool to images instead of signal processing features. Considering that principal components can be visualized as spectrograms, the user can recognize the dominant correlations underlying data. The connection thereby established between the model and the user’s ability to visualize part of the model’s nucleus can be considered a step towards the exploration of interpretations. Moreover, this study provides a quantitative measure of the contribution given by each eigen-spectrogram to a given sample spectrogram. The capability to weigh the contribution given by each feature to a certain prediction is one of the peculiarities of interpretable models.

The AI model was validated by means of two different datasets. The first is the well-known benchmark Case Western Reserve University bearing dataset (CWRU) \cite{0CWRU,Smith2015Rolling}, whereas the second was obtained from noisy signals numerically simulated. For both datasets it was achieved 100\% accuracy performance. A third dataset was obtained by adding noise to numerical signals. Furthermore, the model effectiveness was assessed by state of the art comparisons. 

The main tools which constitute eigen-spectrograms models are certainly established, but their conjunction aimed at spectrogram recognition for IFD has been rather slightly explored. Particularly, the visualization of spectrograms PCA modes lays the ground for interpreting the acquired knowledge. The randomization of the subtended algebra aims at sharpening the model efficiency. This is actually a key factor if the methodology is translated in a big data context \cite{Tian2015Big}. In the analyzed context, the straightforwardness of SVM is beneficial, since it mitigates data overfitting and it limits the computational demand. 

\section{Dataset construction}
\subsection{Experimental dataset: CWRU}
\begin{figure*}[h]
    \centering
    \includegraphics[scale=2.2]{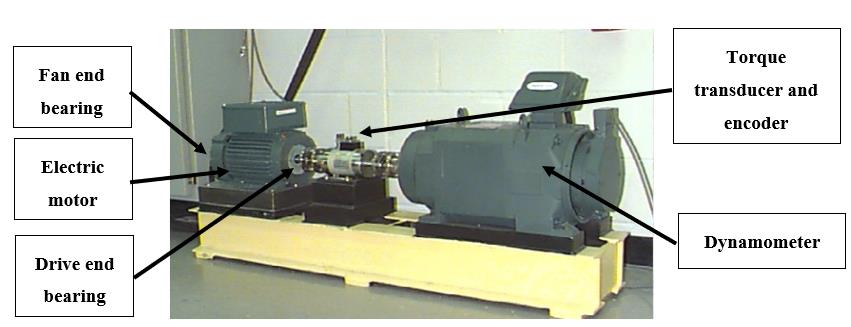}
    \caption{CWRU test bench \cite{Brusa2020Health}}
    \label{fig:2}
\end{figure*}
The CWRU \cite{0CWRU} (Fig.\ref{fig:2}) has become a benchmark dataset for bearing fault diagnosis, despite some limitations \cite{Smith2015Rolling}. The two 2-hp electric motor drives the main shaft. Also, the system is equipped with a torque transducer/encoder, a dynamometer and control electronics. Two test ball bearings are placed on the motor shaft drive end (DE) and fan end (FE). 
\begin{table}[H]
\setlength{\tabcolsep}{4pt}
\renewcommand{\arraystretch}{0.9}

    \centering
    \caption{Bearings characteristic fault frequencies as multiples of shaft speed}
    \label{tab:1}
    \begin{tabular}{c c c c c}
    \toprule
    
    \textbf{Location} & \textbf{Designation} & \textbf{BPFI} & \textbf{BPFO} & \textbf{BSF}\\
   
    \midrule
    
    Drive End & SKF 6205-2RS JEM \footnotemark & 5.415 & 3.585 & 2.357\\
    
    Fan End & SKF 6203-2RS JEM & 4.947 & 3.053 & 1.994\\
    
    Numerical & SKF 22240 CCK/W33 & 11.103 & 7.897 & 2.830\\
    \bottomrule
    \end{tabular}
\end{table}
\footnotetext{NTN-equivalent bearings were used for 0.028 in faults}
Table \ref{tab:1} includes bearings specifications in terms of \lq Ball Passing Frequency on the Inner race\rq (BPFI), \lq Ball Passing Frequency on the Outer race\rq (BPFO) and \lq Ball Spin Frequency\rq (BSF) of which even harmonics (2$\times$BSF) characterize the envelope spectra of damaged rolling elements \cite{Smith2015Rolling}.

Electro-discharge machining (EDM) enabled the introduction of localized faults in bearing elements. In this study, 0.007, 0.014, 0.021 and 0.028 in diameter faults were analyzed for damaged DE balls and inner race. Instead, 0.028 in fault was not analyzed for the DE outer race, since data were not available. Tests were conducted for each damage level by running the motor with powers of 0, 1, 2 and 3 hp, as declared by CWRU \cite{0CWRU}. Shaft speed, assumed to be constant for each test, actually has been variable between 1721 and 1796 rpm. Baseline vibration data were acquired with a sampling frequency $f_s=48$ kHz, whereas fault bearings vibration data were sampled at 12 kHz. Signals were acquired using a 16-channel DAT recorder. Vibration signals coming from the accelerometer set alongside the direction of the gravitational load (6 o’clock direction in \cite{0CWRU,Smith2015Rolling}) were investigated for outer race faults diagnosis.

\subsection{Numerical dataset}
The numerical dataset was constructed by means of a well-established model available in the literature for simulating bearing fault signals \cite{Wang2013enhanced, Wang2011adaptive}. The model of Eq.\ref{Eq:1} and Eq.\ref{Eq:2} assumes that such signals s(t) consists of periodic bursts of exponentially decaying sinusoids (Wang et al., 2013):
\begin{equation}
s(t)=\sum_{j=1}^N A_j h(t-jT)
\label{Eq:1}
\end{equation}
\begin{equation}
h(t)=
\begin{cases}
e^{-\beta t} \sin(2 \pi f_n t)    & t>0 \\
0                              & \text{otherwise}
\end{cases}
\label{Eq:2}
\end{equation}
where $J$ is the number of fault pulses, $A_j$ is the amplitude of the $jth$ impulse, $T$ is the time period corresponding the characteristic fault frequency, $h(t)$ is the impulse function containing the decay parameter $\beta$ and the excited resonance frequency $f_n$. $\beta=1200$ Hz and $f_n=2000$ Hz were selected as simulation parameters. Four levels of a dimensionless $A_j$ were hypothesized. Namely, the mean values of 1, 2, 3, 4 were assigned to $A_j$ by adding also a random oscillating part uniformly varying in the range of $+/-10\%$ the mean $A_j$. The simulated signals are treated as dimensionless. Finally, the resulting signal is passed through an \lq Additive White Gaussian Noise\rq (AWGN) filter to simulate noisy vibration data.
\begin{figure*}[h]
    \centering
    \includegraphics[scale=0.5]{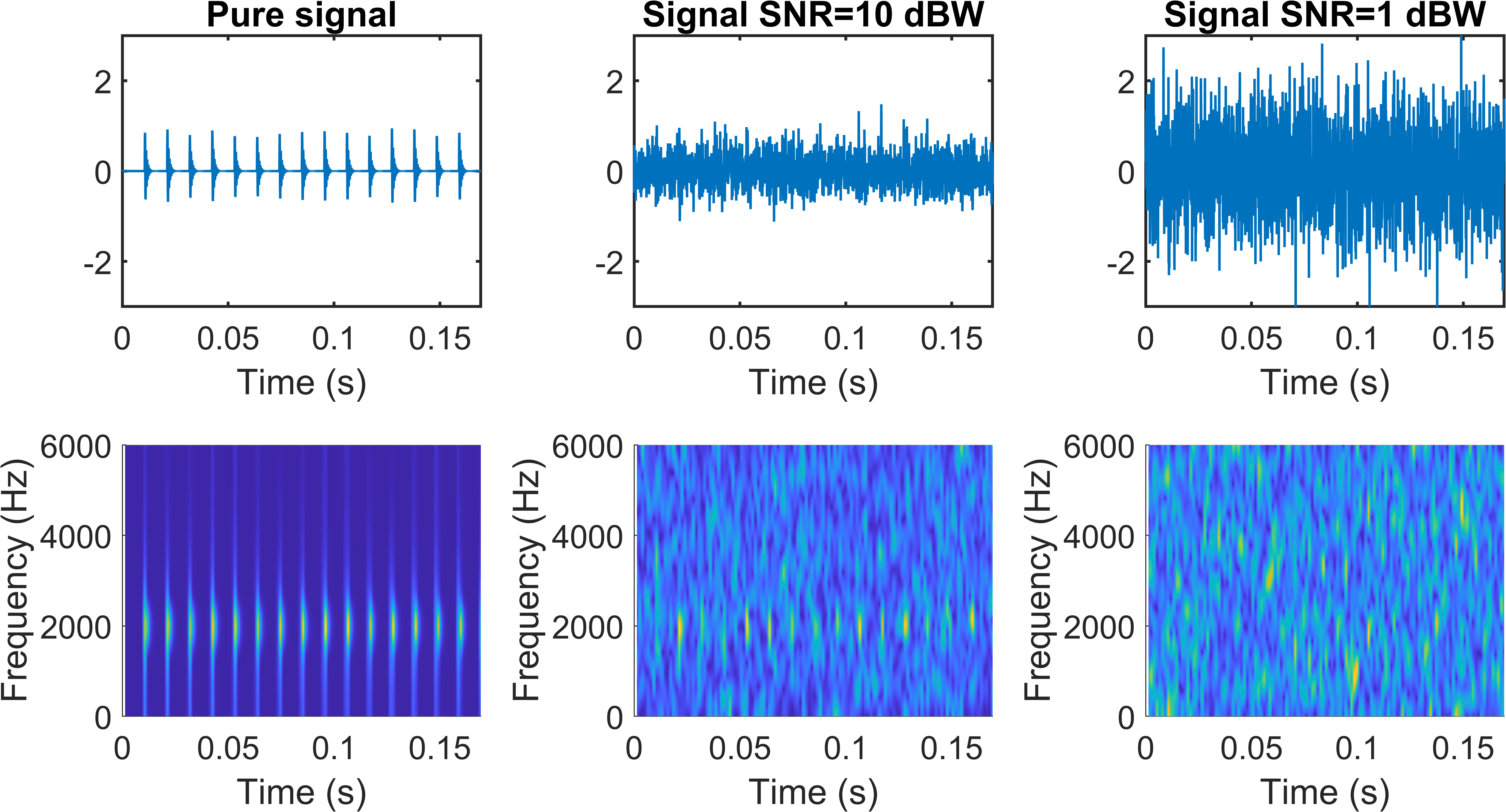}
    \caption{Numerical signal simulated for rolling elements damages and $A_j=1$ with different SNR levels.}
    \label{fig:3}
\end{figure*}
The model considered \lq Signal-to-Noise Ratio\rq (SNR) levels of 10 dBW and 1 dBW (Fig.\ref{fig:3}). The numerical signals were sampled at 12 kHz. Inner race, outer race and rolling elements damages were simulated for the bearing SKF 22240 CCK/W33 running at 1000 rpm. Table \ref{tab:1} reports the characteristic frequencies computed for the analyzed bearing.

\subsection{Spectrogram images}
The time-frequency representations (Fig.\ref{fig:3}) of the signals were constructed by means of the \lq Short Time Fourier Transform\rq (STFT) applied with 32 samples Hamming window and 50\% overlap. This set of parameters enabled a good trade-off to achieve adequate frequency and time resolutions.

To take advantage of a statistically significant number of samples for the training process, data augmentation was carried out on vibration signals. Actually, thousands of samples define the current state of the art for IFD of REB. As already highlighted, this is an inherent limit of existing benchmark datasets, outlined by the capability of catching chunks physically meaningful. Excessively small chunks could indeed contain few fault pulses. For this reason, the limit for data segmentation was fixed to 2048 samples. In this way, each chunk contained a minimum of 16 fault impulses (2$\times$BSF in the numerical dataset).

\begin{figure*}[h]
    \centering
    \includegraphics[scale=0.55]{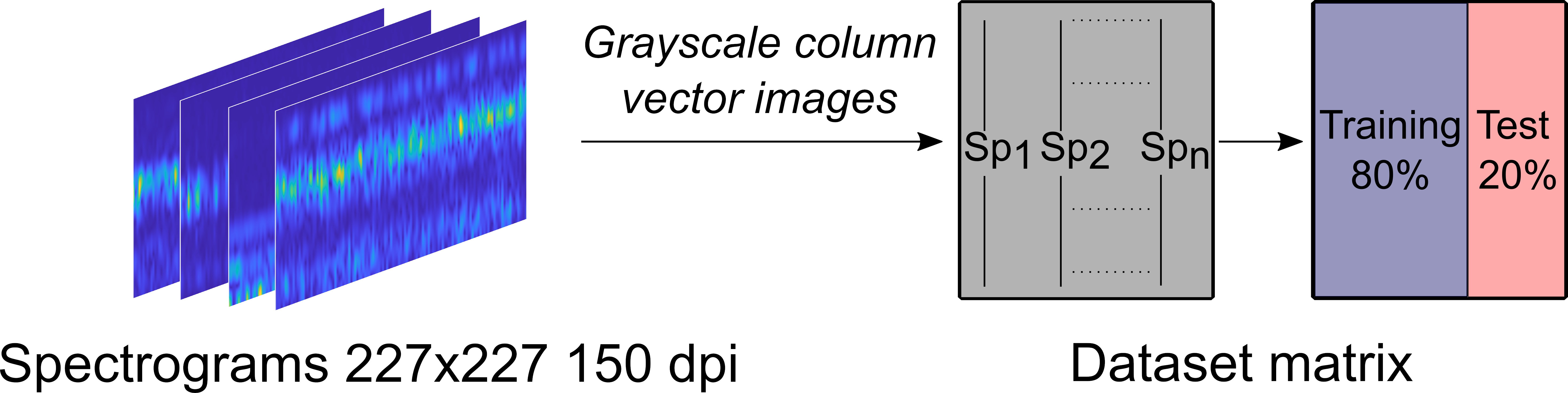}
    \caption{Dataset construction.}
    \label{fig:4}
\end{figure*}
Finally, the spectrograms were converted in 227$\times$227 resolution images, commonly adopted for CNNs. Each image was turned in a grayscale picture and reshaped in a column vector to construct the dataset matrix (Fig.\ref{fig:4}).

\begin{table*}[h]
\setlength{\tabcolsep}{4pt}
\renewcommand{\arraystretch}{0.9}

    \centering
    \caption{Class labels, training samples and test samples for datasets}
    \label{tab:2}
    \begin{tabular}{c c c c c c c c c c c c c}
    
    \toprule
    \multicolumn{13}{c}{CWRU dataset. 3423 images 150 dpi, 227$\times$227}\\
    
    \midrule
    \textbf{Class} & B007 & B014 & B021 & B028 & IR007 & IR014 & IR021 & IR028 & OR007 & OR014 & OR021 & Normal\\
    
    \textbf{Train} & 189 & 189 & 189 & 188 & 190 & 189 & 188 & 187 & 189 & 189 & 189 & 663\\
    
    \textbf{Test} & 47 & 47 & 47 & 47 & 47 & 47 & 48 & 48 & 47 & 47 & 47 & 165\\
    
    \toprule
    \multicolumn{13}{c}{Nuerical dataset. 1800 images 150 dpi, 227$\times$227}\\
    
    \midrule
    \textbf{Class} & B1 & B2 & B3 & B4 & IR1 & IR2 & IR3 & IR4 & OR1 & OR2 & OR3 & OR4\\
    
    \textbf{Train} & 120 & 120 & 120 & 120 & 120 & 120 & 120 & 120 & 120 & 120 & 120 & 120\\
    
    \textbf{Test} & 30 & 30 & 30 & 30 & 30 & 30 & 30 & 30 & 30 & 30 & 30 & 30\\
    \bottomrule
    \end{tabular}
\end{table*}

The CWRU dataset consists of 3423 images, whereas 1800 images are extracted from the numerical signals. Images were labelled by following an alphanumeric codification. The first part of the code indicates with B, IR and OR respectively rolling elements, inner race and outer race faults. The second part expresses the damage severity for the CWRU dataset and the amplitude $A_j$ for the numerical set. Datasets were randomly partitioned for training (80\%) and test (20\%) as showed in Table \ref{tab:2}. For each CWRU class, data comes from 0, 1, 2 and 3 hp vibration signals. Overall, CWRU data consisted of 2739 training samples and 684 test samples, whereas the numerical dataset consisted of 1440 training samples and 360 test samples.

\section{Feature extraction via rSVD: Eigen-spectrograms}
In the imminent big data perspective, it is quite common to assume that increasingly large amount of information coming from IoT systems still keeps the dominant engineering message in inherent low-rank structures \cite{Brunton2019Data}. In this sense, the deterministic matrix decomposition starts to be computationally hard when facing large datasets.

Randomized algebra offers a mean to perform such data mining tasks through random sampling, which sharply reduce the computational effort. Over the past decades, these theories have found consistent mathematical background \cite{Halko2011Finding,Martinsson2011randomized,Sarlos2006Improved}. In this work, the rSVD algorithm \cite{Erichson2019Randomized} by Halko et al. \cite{Halko2011Finding}, freshly resumed by \cite{Brunton2019Data}, is applied to traditional PCA \cite{Pearson1901On,Zhang2020Machine-learning-based}. In the following, bold capital letters refer to matrices, whereas bold lowercase letters indicate vectors.

\subsection{rSVD for principal component analysis}
The training set matrix $\bm{X} \in \mathbb{R}^{n \times m}$  containing $n$ rows (227$\times$227 pixels) and $m$ columns (training samples) is at first used to compute the mean column vector $\bm{s}=\frac{1}{m}\sum_{j=1}^{m}\bm{X_{ij}}$, corresponding to the mean spectrogram image. 
\begin{equation}
\bm{B}=\bm{X}-\bm{s}
\begin{bmatrix}
1 & \dotsm & 1
\end{bmatrix}
=\bm{X}-\bar{\bm{X}}
\label{Eq:3}
\end{equation}
\begin{equation}
\bm{Z}=\bm{B}\bm{P}
\label{Eq:4}
\end{equation}
The mean centered training dataset $\bm{B}$ expressed by Eq.\ref{Eq:3} is sampled through the random projection matrix $\bm{P} \in \mathbb{R}^{m \times r}$  to obtain the matrix $\bm{Z} \in \mathbb{R}^{n \times r}$, which approximates the column space of $\bm{B}$ (Eq.\ref{Eq:4}). $r$ is the target rank and the elements of the matrix $\bm{P}$ are drawn from the standard normal distribution. Essentially, the randomization process lies in the matrix $\bm{P}$. 

The core idea is that a random projection of $\bm{B}$ still keeps the important features contained in the original data. Then, a low-rank orthonormal base of $\bm{B}$ is extracted by computing the QR decomposition $\bm{Z=QR}$. Consequently, $\bm{B}$ is projected in the low dimensional space defined by the matrix $\bm{Q}$:
\begin{equation}
\bm{Y}=\bm{Q^T}\bm{B}
\label{Eq:5}
\end{equation}
and the SVD $\bm{Y=U_Y\Sigma V^T}$ is computed.
The SVD matrix decomposition results in the left singular vectors (columns of $\bm{U_Y}$), the singular values (on the diagonal of $\bm{\Sigma}$) and the right singular vectors (columns of $\bm{V}$). Finally, the left singular vectors $\bm{U}$ of the matrix $\bm{B}$ can be reconstructed by projecting $\bm{U_Y}$ in the original space $\bm{U=QU_Y}$. 
\begin{equation}
(\bm{BB^T})\bm{U}=\bm{U\Sigma^2}
\label{Eq:6}
\end{equation}
From a PCA standpoint, it is possible to demonstrate that the SVD of the mean centered data $\bm{B}$ solves the eigenproblem of Eq.\ref{Eq:6} associated to the matrix $\bm{C=BB^T}$. This matrix denotes, in this particular application, a measure of pixels’ correlations. Thus, the directions capturing the variance linked to the singular values $\bm{\Sigma_{ii}}$ are hierarchically defined and ordered as column vectors of $\bm{U}$ (principal components).

\subsection{Eigen-spectrograms}
As anticipated in the introductory paragraphs, the left singular vectors stored in the columns of U were denoted as \emph{eigen-spectrograms}, inspired by similar research conducted in the field of facial recognition \cite{Belhumeur1997Eigenfaces,Brunton2019Data,Kirby1990Application,Sirovich1987Low-dimensional,Turk1991Eigenfaces}. In this work, eigen-spectrograms encapsulate the compelling aspect of offering interpretations for PCA results and, as aftermath, of the IFD model. 
\begin{figure}[ht]
    \centering
    \includegraphics[scale=0.45]{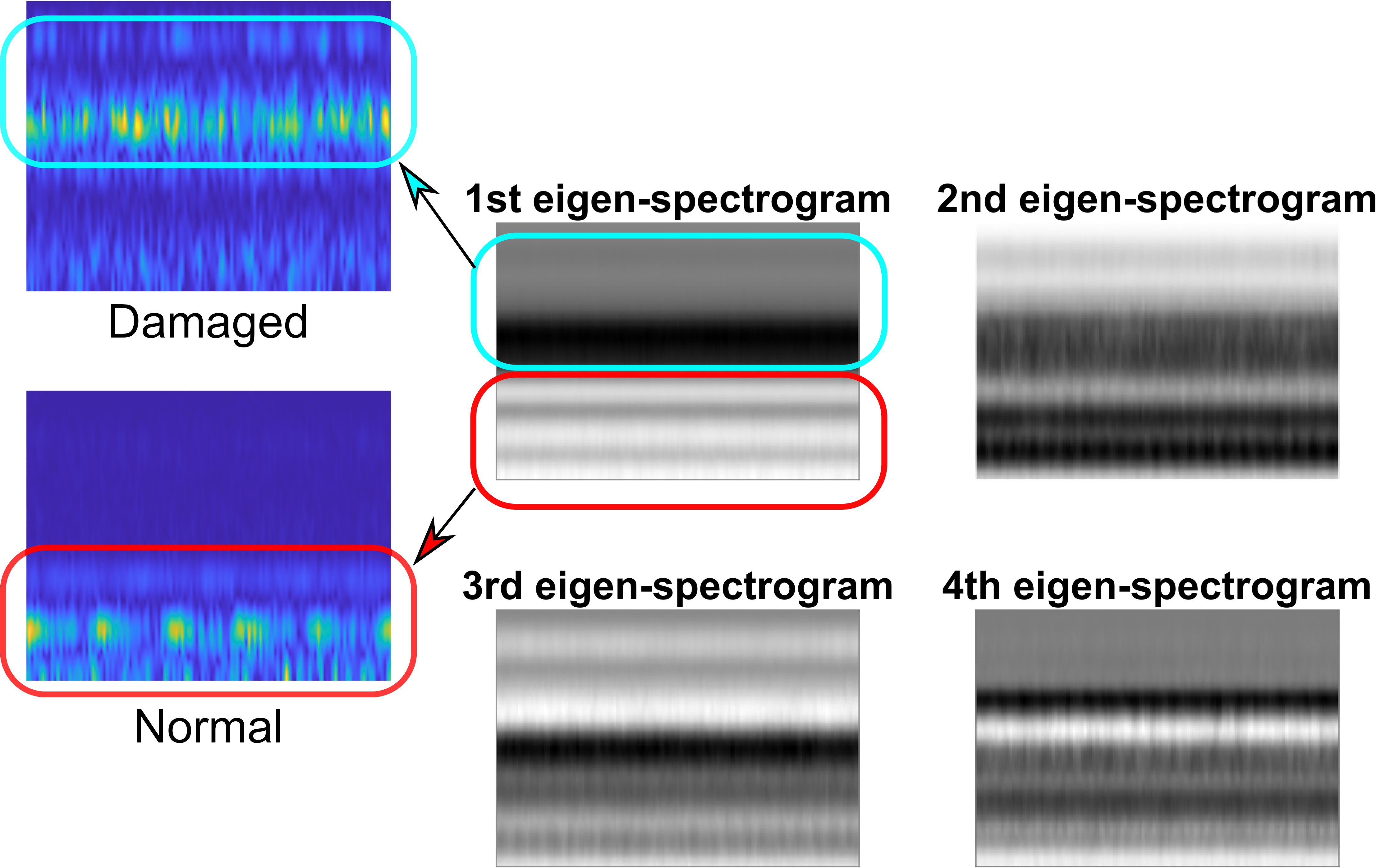}
    \caption{Eigen-spectrograms visualized as reshaped grayscale images. CWRU dataset.}
    \label{fig:5}
\end{figure}
\begin{figure}[ht]
    \centering
    \includegraphics[scale=0.6]{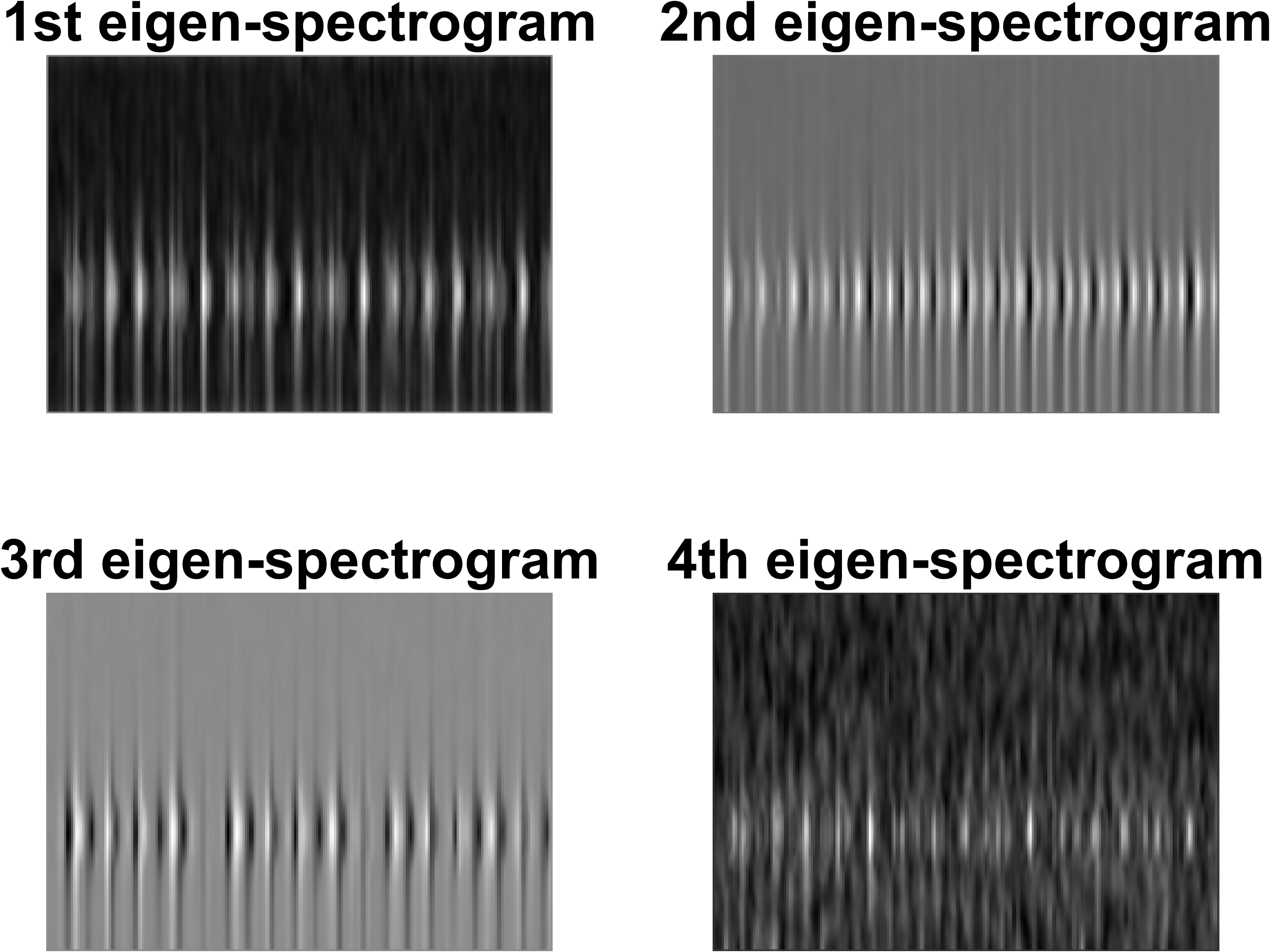}
    \caption{Eigen-spectrograms visualized as reshaped grayscale images. Numerical data with 10 dBW SNR.}
    \label{fig:6}
\end{figure}
Indeed, these vectors, besides being related to principal components directions, can be reshaped as 227$\times$227 grayscale images (Fig.\ref{fig:5} and Fig.\ref{fig:6}). In these representations, the highest grayscale values are assigned to white pixels, whereas black pixels stand for the lowest values.Hence, the lighter zones emphasize the spectrogram portions which mostly contribute to explain the dataset variance associated to a certain principal component (Eq.\ref{Eq:6}). For example, the first eigen-spectrogram is able to catch the regions where CWRU spectrograms mostly differ (Fig.\ref{fig:5}) in terms of fault and healthy state.

The AI classifier was constructed by holding the first four principal components. Actually adding other principal components, although maintaining maximum accuracy, would vainly raise the computational effort for the training phase. With the intention of correctly describing the first four principal components, random projections of the rSVD algorithm were computed by using a rank $r=110$, one order of magnitude below the original column space dimensionality (thousands-samples).
\begin{figure*}[h]
    \centering
    \includegraphics[scale=0.7]{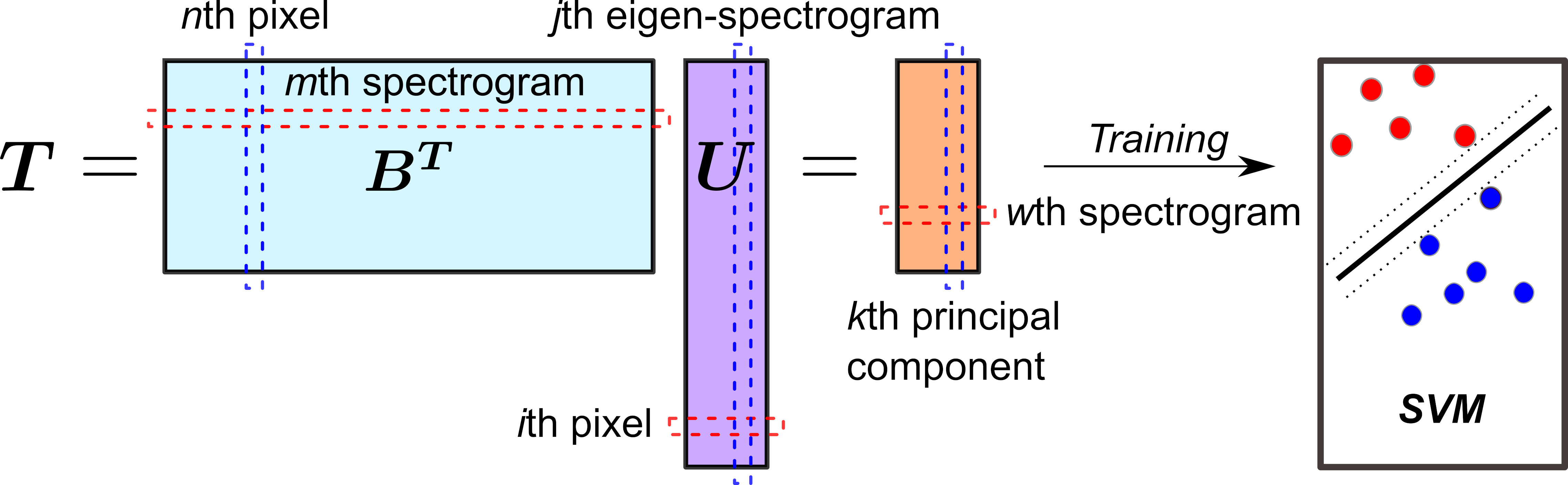}
    \caption{Dataset projection in the eigen-spectrograms space.}
    \label{fig:7}
\end{figure*}
Finally, the training set is projected in the principal component subspace defined by the four eigen-spectrograms with the transformation $\bm{T=B^TU}$ (Fig.\ref{fig:7}). In other words, each spectrogram is pixel-by-pixel weighted giving more importance to lighter eigen-spectrograms regions.

The $jth$ feature $F_{ij}= \langle \bm{b_i,u_j} \rangle $ of the $ith$ spectrogram is therefore given by the scalar product between the $ith$ spectrogram $\bm{b_i}$ and the $jth$ eigen-spectrogram $\bm{u_j}$. 
\begin{figure*}[h]
    \centering
    \includegraphics[scale=0.4]{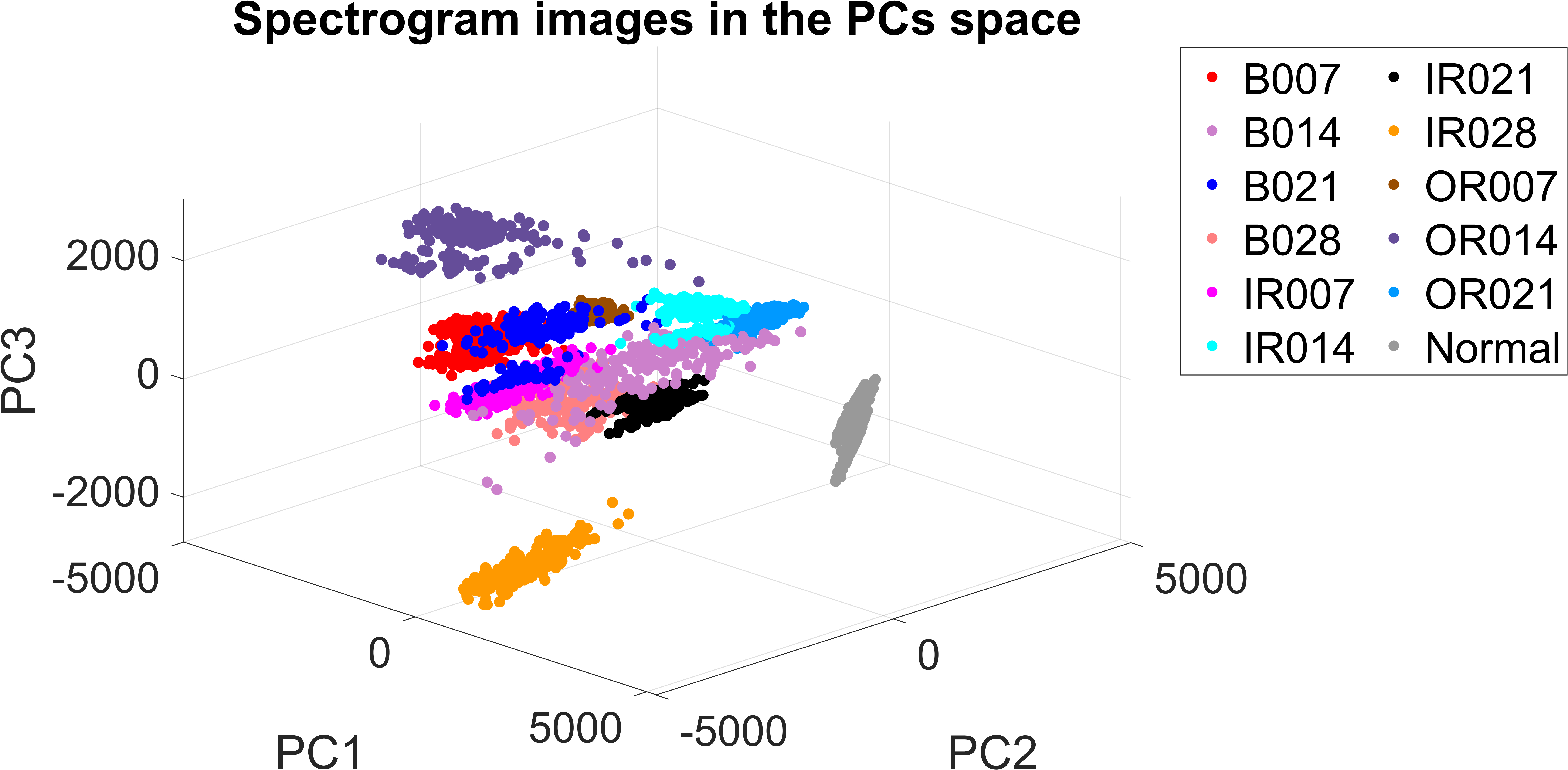}
    \caption{Spectrograms in principal components coordinates. CWRU dataset.}
    \label{fig:8}
\end{figure*}
Then, each row of the matrix $\bm{T}$ describes the corresponding spectrogram in the low-dimensional space defined by the first four eigen-spectrograms (Fig.\ref{fig:8}).
The complete SVD decomposition would bring to square orthogonal $\bm{U}$ matrices and each spectrogram $\bm{b_i}$ could be expressed by the linear combination of eigen-spectrograms $\bm{b_i}=\bm{Ut_i^T}$, where $\bm{t_i}$ is the $ith$ row of the matrix $\bm{T}$. On the other hand, the low-rank approximation enables to reconstruct spectrograms by using only some columns of the eigen-spectrogram matrix: $\bm{b_i}\simeq \bm{Ut_i^T}$. In short, the features $F_{ij}$ weigh the contribution given by the $jth$ eigen-spectrogram to the $ith$ spectrogram. In this sense, eigen-spectrograms showed to combine differently for distinct machine health states (Fig.\ref{fig:8}). Moreover, thanks to the use of images, the principal component subspace can be displayed (Fig.\ref{fig:5}).

\subsection{Interpretation coefficients}
Noticeably, the expression $\bm{b_i}\simeq\bm{Ut_i^T}$ is equivalent to  $\bm{b_i}\simeq\sum_{j=1}^{4} F_{ij} \bm{u_j}$, which better emphasizes how the eigen-spectrograms (columns of $\bm{U}$) combine to approximate the image $\bm{b_i}$. Additionally, it is possible to obtain the dimensionless scalar factor $\Gamma_i$ of Eq. (\ref{Eq:7}) by multiplying the previous expression by $\bm{b_i^T}$.
\begin{equation}
\Gamma_i=\frac{\sum_{j=1}^{4} F_{ij} \bm{b_i^T u_j}}{\bm{b_i^T b_i}}
\label{Eq:7}
\end{equation}

This mathematical formulation offers the advantage of evaluating the importance that each feature has in reconstructing a given spectrogram. Indeed, each term of the summation can be interpreted as the relative contribution given by the $jth$ eigen-spectrogram to the $ith$ spectrogram. Clearly, a factor $\Gamma_i=1$ would mean that all the principal components were taken into account. Since the model considers four principal components, only a portion of the dataset variance can be explained and $\Gamma_i$ will be $\Gamma_i<1$.   Nevertheless, it is possible to introduce the interpretation coefficients $\theta_j$ reported in Eq.(\ref{Eq:8}) by normalizing Eq.(\ref{Eq:7}) with respect to $\Gamma_i$. In such a case, the summation is not only dimensionless but also unitary. 

\begin{equation}
\theta_j=\frac{F_{ij}\bm{b_i^T u_j}}{\Gamma_i \bm{b_i^T b_i}}=\frac{F_{ij}^2}{\Gamma_i \bm{b_i^T b_i}},
\qquad \sum_{j=1}^{4} \theta_j=1
\label{Eq:8}
\end{equation}

The interpretation coefficients $\theta_j$ offer a means to evaluate how much the $jth$ eigen-spectrogram is involved in the \lq explainable\rq  part of the $ith$ spectrogram, represented by the term $\Gamma_i \bm{b_i^T b_i}$. Eq. (\ref{Eq:8}) shows that the features $F_{ij}$ can be interpreted as absolute contributions, whereas $\theta_{j}$ stand for relative contributions. This supports the idea that each feature is tightly connected to the weight that each eigen-spectrogram hold in representing a specific image.
\begin{figure}[ht]
    \centering
    \includegraphics[scale=0.6]{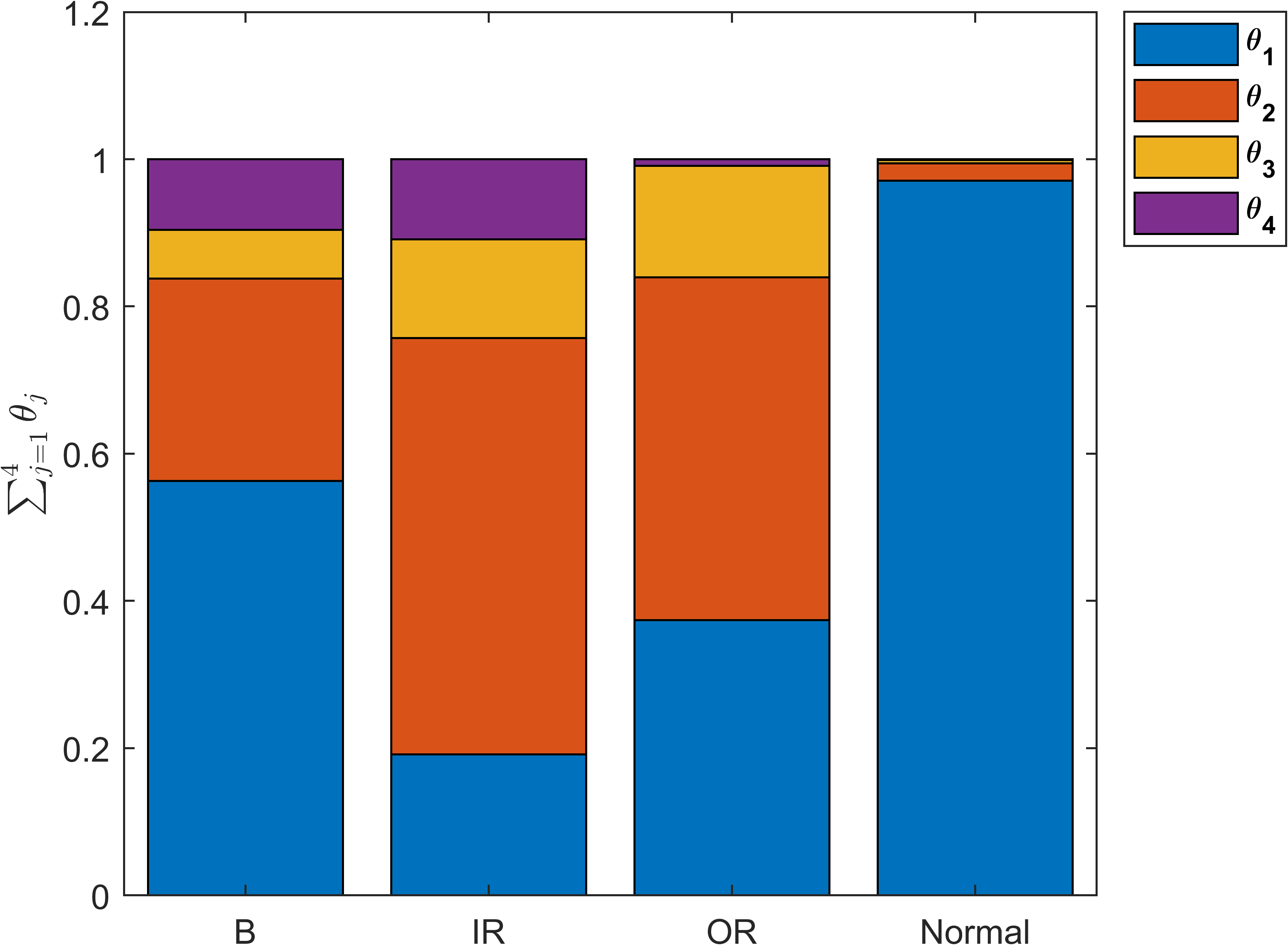}
    \caption{Mean interpretation coefficients.}
    \label{fig:9}
\end{figure}
Fig.\ref{fig:9} shows the mean interpretation coefficients computed for different damage types using 300 random samples of each class. Remarkably, the first eigen-spectrogram mostly explains normal data, as inferred from Fig.\ref{fig:5}. Conversely, damages can be explained by considering relevant contributions by the second feature, embodying the second eigen-spectrogram. 
Thereby, this formulation provides a tool for interpreting the feature space, and it shows how data are decomposed through such a space. In this case, the potentials to glimpse reasons subtended to the decisions of a data-driven model define model’s interpretability.

\subsection{Kernelized multiclass SVM}
The three-dimensional portrayal provided in Fig.\ref{fig:8} gives an insight into the rows of the matrix $\bm{T}$ which include four elements. Thus, each element corresponds to a specific coordinate in the principal component space, and it is employed as a feature for the training of the machine learning classifier.

Already by eye inspection of data in the eigen-spectrogram space (Fig.\ref{fig:8}), spectrograms belonging to different classes appeared quite discernable. Then, it is reasonable to assume the existence of a link underlying eigen-spectrograms scalar products $F_{ij}= \langle \bm{b_i,u_j} \rangle$ and machine health state. The boundaries that enclose eigen-spectrograms combinations $F_{ij}$ associable to the same health state are sought by means of a SVM classifier.

SVM \cite{Brunton2019Data, Cortes1995Support-vector,Widodo2007Support} is one of the most exploited machine learning algorithms for decision boundaries tracking. In its original form, it enables binary classification of linearly separable data. However, thanks to proper mathematical treatment it can be easily applied to multiclass non-linear problems thanks to kernel methods.

Namely, a second order polynomial kernel was employed in this study, as it provided best outcomes. Given that SVM is constructed for binary classification, several strategies have also evolved to deal with multiclass problems. In this research, an \lq Error-Correcting Output Codes\rq (ECOC) \cite{Escalera2009} model was employed to label data coming from twelve different classes (Table \ref{tab:2}). Each one of the ECOC binary classifiers makes use of a cost parameter $C=1.0$. The optimization of hyperparameters is beyond the purposes of this study, however grid search algorithms could be implemented for fine-tuning the fault diagnosis models. 

\section{Results and discussion}
The proposed methodology leverages on the possibility of interpreting feature space, constructed via image processing. 
\begin{table}
\setlength{\tabcolsep}{4pt}
\renewcommand{\arraystretch}{0.9}

    \centering
    \caption{Machine Learning classifier for IFD via spectrogram image processing. Specifications.}
    \label{tab:3}
    \begin{tabular}{l c}
    
    \toprule
    
    \textbf{Features} & \makecell{Dataset coordinates\\in the eigen-spectrograms space\\(computed by means of rSVD)}\\ 
    
    \midrule
    
    \textbf{Number of features} & 4\\
    
    \midrule
    
    \textbf{Classifier} & SVM\\
    
    \midrule
    
    \textbf{Kernel} & Quadratic, $K(\bm{x_j},\bm{x})=(\bm{x_j}\cdot\bm{x}+1)^2$\\
    
    \midrule
    
    \textbf{Multiclass model} & \makecell{ECOC\\(Error-Correcting Output Codes)}\\
    
    \midrule
    
    \textbf{Validation} & 5-fold cross-validation\\
    
    \bottomrule

    \end{tabular}
    
\end{table}

\begin{table}
\setlength{\tabcolsep}{2pt}
\renewcommand{\arraystretch}{0.7}

    \centering
    \caption{Model accuracy.}
    \label{tab:4}
    \begin{tabular}{l c c c}
    \toprule
    
    \textbf{Dataset} & CWRU & \makecell{Numerical\\SNR 10 dBW} & \makecell{Numerical\\SNR 1 dBW}\\
    \midrule
    \textbf{Training (\%)} & 99.89\% & 100\% & 98.82\%\\
    \midrule
    \textbf{Cross-validation\footnotemark (\%)} & 99.85\% & 100\% & 97.78\%\\
    \midrule
    \textbf{Test (\%)} & 100\% & 100\% & 98.89\%\\
    
    \bottomrule
    \end{tabular}
\end{table}
\footnotetext{Cross-validation accuracy refers to the mean accuracy of the 5 models}
Table \ref{tab:3} summarizes the main features of the spectrograms classifier built for fault detection of REB, whereas Table \ref{tab:4} reports the accuracies obtained for the different datasets. The highest possible accuracy was reached for the CWRU’s data and for the numerical dataset with 10 dBW SNR, whereas the accuracy moderately deviates from 100\% for 1 dBW numerical signals.

Cross-validation accuracy resulted from the average accuracy of the 5 models that were obtained from 5-fold random partitions of the training set. The similarity between metrics achieved for the whole training sets and for the cross-validated models suggested that training data contained a meaningful number of samples and, no less, overfitting was unlikely to occur. Therefore, the validation phase suggested that it was worth testing the machine learning model under never seen data. Accuracies outlining training and test data are indeed very similar. It may therefore be reasonable to suppose that the model efficiently generalizes the learned knowledge, which in this case is portrayed by the non-linear SVM boundaries.
\begin{figure}[ht]
    \centering
    \includegraphics[scale=0.75]{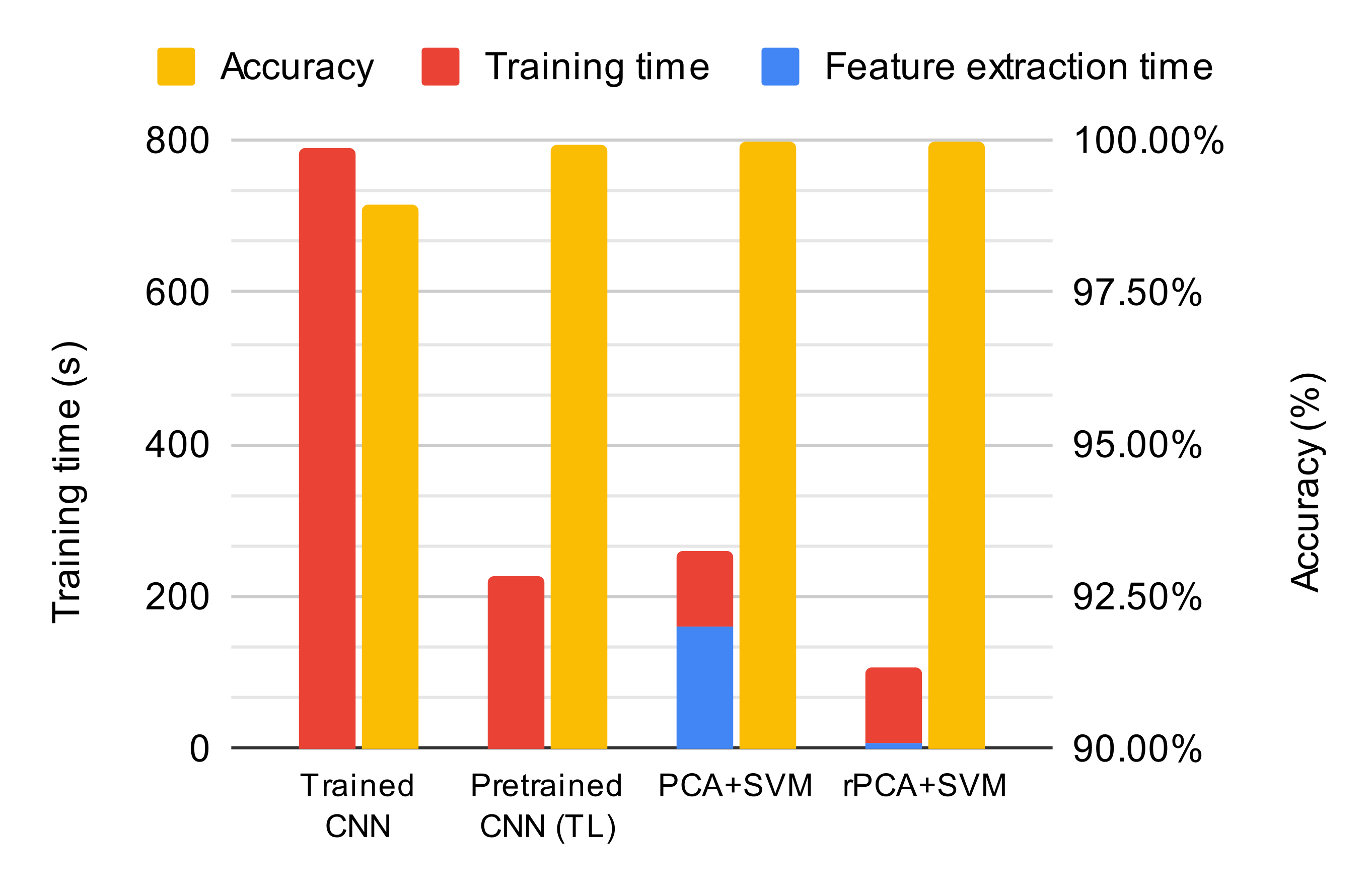}
    \caption{Model performances. Comparison with literature models.}
    \label{fig:10}
\end{figure}

\begin{table*}
\setlength{\tabcolsep}{3pt}
\renewcommand{\arraystretch}{1.8}

    \centering
    \caption{Accuracy on the CWRU. Comparison with literature models.}
    \label{tab:5}
    \begin{tabular}{l c c c}
    \toprule
    
    \textbf{Model} & \textbf{Features} & \textbf{Feature extraction} & \textbf{Accuracy}\\
    \midrule
    
    \makecell[l]{Semisupervised\\self-organizing map \cite{Li2013}} & \makecell[c]{Time domain + bearing\\ characteristic frequencies} & Manual & 95.80\%\\
    
    \midrule
    
    SVM \cite{Zhang2015} & \makecell[c]{Permutation entropy of IMFs\\decomposed by EEMD} & Manual & 97.91\%\\ 
    
    \midrule
    
    Neural Network \cite{Lei2016} & Unsupervised feature learning & Automated & 99.66\%\\    
    
    \midrule
    
    CNN \cite{Ding2017} & Convolutional feature extractor & Automated & 99.80\%\\   
    
    \midrule
    
    Recurrent Neural Network (RNN) \cite{Zhao2018} & \makecell[c]{Local features +\\bidirectional GRU network} & Hybrid & 99.60\%\\   
 
    \midrule
    
    Transfer learning \cite{Shao2019Highly} & CNN pre-trained for images & Automated & 99.95\%\\ 
    
    \midrule
    
    Transfer learning \cite{Brusa2021Deep} & CNN pre-trained for audios & Automated & 100\%\\ 
    
    \midrule
    
    Present work & Eigen-spectrograms & Manual & 100\%\\  
    \bottomrule
    \end{tabular}
\end{table*}

Fig.\ref{fig:10} shows state of the art results \cite{Shao2019Highly} for the CWRU dataset. The computational effort is dramatically reduced with respect to CNNs trained from scratch under wavelet time-frequency images \cite{Shao2019Highly} and accuracy is slightly improved. Under a computational perspective SVM is extremely light given that it manages a very low number of parameters as compared with CNNs. Further, it can be assumed that this aspect counteracts overfitting in datasets, which include thousands of samples (as it is in the case of REB) rather than millions (as in the case of images sets).

TL strategies, instead, achieve performances comparable to standard PCA+SVM. Indeed, PCA for feature extraction acts as a bottleneck of the whole training process. Remarkably, rPCA applied via rSVD substantially reduces the time for the features extraction, while still keeping improved accuracy. In this case, the computing resources are engaged by the only SVM, which confirms to be a very low-impact algorithm. Halving training time with respect to pretrained models and standard PCA can be regarded as an overwhelming advantage, especially under a big data perspective. Nevertheless, in such a context, the applicability of the methodology needs further investigations. 

Accuracies available in the literature for the CWRU dataset are reported in Table \ref{tab:5}. Also, the features type and the feature extraction method are included in Table \ref{tab:5}. It is possible to observe that eigen-spectrograms diagnosis achieves state-of-the-art performances.

\section{Conclusion}
This research was encouraged by the leading question which has driven authors’ investigations on the base of the current literature: does statistical learning  provides yet profitable opportunities for exploring REB fault diagnosis in the forthcoming big data decades? It is concluded that:
\begin{itemize}
    \item spectrograms image processing by means of machine learning offers remarkable performances for bearings fault diagnosis, since SVM-based classifiers comfortably apply to spectrograms recognition;
    \item rPCA is able to catch meaningful as long as interpretable features;
    \item it is introduced the concept of eigen-spectrogram, embodying feature space;
    \item the eigen-spectrograms feature space can be interpreted using proper coefficients;
    \item randomized algebra shows to be a promising as much as fascinating engineering device to extract low-rank structures underlying bearings datasets. By virtue of its computational benefits, it is proposed as an opportunity to approach new reading keys of statistical learning theories in next IoT datasets.
\end{itemize}

This work shows that eigen-spectrograms efficiently capture, with restrained human assistance, inherent structures in vibration data coming from a specific machine. Also, model interpretations can be examined since spectrograms regions mostly contributing to dataset variance are hierarchically displayable by employing eigen-spectrograms. Although the model effectively generalizes knowledge from the training to the test set, the generalization abilities with respect to unseen working conditions should be further explored. The applicability of the proposed model to larger and harder standardized dataset calls for new experimental validations and, also, RUL assessments could be evaluated. Clearly, the paucity of bearing faults data in industrial contexts limits the applicability of the methodology. Future work will investigate the applicability of the methodology to different datasets. May eigen-spectrograms be further generalized? Are they able to catch low-rank structures in test data coming from a completely unseen machine? How to build eigen-spectrograms providing this peculiarity? These attractive questions lay the ground for future research.

\section*{Data availability statement}
The data that support the findings of this study are available at \url{https://engineering.case.edu/bearingdatacenter}

\section*{Declaration of conflicting interests}
The authors report there are no competing interests to declare.


\normalsize
\bibliography{ms}

\begin{thebibliography}{10}

\bibitem{Mohanty2014Machinery}
A.~R. Mohanty, {\em Machinery condition monitoring: Principles and practices}.
\newblock CRC Press, 2014.

\bibitem{Randall2011Vibration-based}
R.~B. Randall, {\em Vibration-based condition monitoring: industrial, aerospace
  and automotive applications}.
\newblock John Wiley \& Sons, 2011.
\newblock Book on VBCM - Matlab pages on predictive maintenance are based on
  this book.

\bibitem{Brusa2020Design}
E.~Brusa, ``Design of a kinematic vibration energy harvester for a smart
  bearing with piezoelectric/magnetic coupling,'' {\em Mechanics of Advanced
  Materials and Structures}, vol.~27, no.~15, pp.~1322--1330, 2020.

\bibitem{Brusa2021Thermal}
E.~Brusa, M.~D. Vedova, L.~Giorio, and P.~Maggiore, ``Thermal condition
  monitoring of large smart bearing through fiber optic sensors,'' {\em
  Mechanics of Advanced Materials and Structures}, vol.~28, no.~11,
  pp.~1187--1193, 2021.

\bibitem{Gougam2021Fault}
F.~Gougam, R.~Chemseddine, D.~Benazzouz, K.~Benaggoune, and N.~Zerhouni,
  ``Fault prognostics of rolling element bearing based on feature extraction
  and supervised machine learning: Application to shaft wind turbine gearbox
  using vibration signal,'' {\em Proceedings of the Institution of Mechanical
  Engineers, Part C: Journal of Mechanical Engineering Science}, vol.~235,
  no.~20, pp.~5186--5197, 2021.

\bibitem{Wang2018Prognostics}
D.~Wang, K.-L. Tsui, and Q.~Miao, ``Prognostics and health management: A review
  of vibration based bearing and gear health indicators,'' {\em IEEE Access},
  vol.~6, pp.~665--676, 2018.

\bibitem{McFadden1984Vibration}
P.~McFadden and J.~Smith, ``Vibration monitoring of rolling element bearings by
  the high-frequency resonance technique — a review,'' {\em Tribology
  International}, vol.~17, pp.~3--10, 2 1984.
\newblock This Paper (1984) introduce the HFRT later called Envelope.

\bibitem{Abboud2019Advanced}
D.~Abboud, M.~Elbadaoui, W.~A. Smith, and R.~B. Randall, ``Advanced bearing
  diagnostics: A comparative study of two powerful approaches,'' {\em
  Mechanical Systems and Signal Processing}, vol.~114, pp.~604--627, 2019.

\bibitem{Brusa2020Health}
E.~Brusa, F.~Bruzzone, C.~Delprete, L.~G. Di~Maggio, and C.~Rosso, ``Health
  indicators construction for damage level assessment in bearing diagnostics: A
  proposal of an energetic approach based on envelope analysis,'' {\em Applied
  Sciences (Switzerland)}, vol.~10, no.~22, pp.~1--24, 2020.

\bibitem{Delprete2005Rolling}
C.~Delprete, M.~Milanesio, and C.~Rosso, ``Rolling bearings monitoring and
  damage detection methodology,'' {\em Applied Mechanics and Materials},
  vol.~3-4, pp.~293--302, 2005.

\bibitem{Delprete2020Bearing}
C.~Delprete, E.~Brusa, C.~Rosso, and F.~Bruzzone, ``Bearing health monitoring
  based on the orthogonal empirical mode decomposition,'' {\em Shock and
  Vibration}, vol.~2020, 2020.

\bibitem{Randall2011Rolling}
R.~B. Randall and J.~Antoni, ``Rolling element bearing diagnostics-a
  tutorial,'' {\em Mechanical Systems and Signal Processing}, vol.~25, no.~2,
  pp.~485--520, 2011.
\newblock This paper is a summary of Randall's book.

\bibitem{Sawalhi2008Semi-automated}
N.~Sawalhi and R.~B. Randall, ``Semi-automated bearing diagnostic-three case
  studies,'' {\em Non Destrcutive Testing Australia}, vol.~45, no.~2, p.~59,
  2008.

\bibitem{Zhong2022Fault}
X.~Zhong, Q.~Mei, X.~Gao, and T.~Huang, ``Fault diagnosis of rolling bearings
  based on improved direct fast iterative filtering and spectral amplitude
  modulation,'' {\em Proceedings of the Institution of Mechanical Engineers,
  Part C: Journal of Mechanical Engineering Science}, vol.~0, no.~0,
  p.~095440622110556, 2022.

\bibitem{Abboud2017Envelope}
D.~Abboud, J.~Antoni, S.~Sieg-Zieba, and M.~Eltabach, ``Envelope analysis of
  rotating machine vibrations in variable speed conditions: A comprehensive
  treatment,'' {\em Mechanical Systems and Signal Processing}, vol.~84,
  pp.~200--226, 2017.

\bibitem{Daga2021Fast}
A.~P. Daga, A.~Fasana, L.~Garibaldi, S.~Marchesiello, and A.~Moshrefzadeh,
  ``Fast computation of the autogram for the detection of transient faults,''
  (Cham), pp.~469--479, Springer International Publishing, 2021.

\bibitem{Moshrefzadeh2018Autogram:}
A.~Moshrefzadeh and A.~Fasana, ``The autogram: An effective approach for
  selecting the optimal demodulation band in rolling element bearings
  diagnosis,'' {\em Mechanical Systems and Signal Processing}, 2018.

\bibitem{RohaniBastami2020Rolling}
A.~Rohani~Bastami and A.~Bashari, ``Rolling element bearing diagnosis using
  spectral kurtosis based on optimized impulse response wavelet,'' {\em
  JVC/Journal of Vibration and Control}, vol.~26, pp.~175--185, 2 2020.

\bibitem{Smith2019Optimal}
W.~A. Smith, P.~Borghesani, Q.~Ni, K.~Wang, and Z.~Peng, ``Optimal
  demodulation-band selection for envelope-based diagnostics: A comparative
  study of traditional and novel tools,'' {\em Mechanical Systems and Signal
  Processing}, vol.~134, p.~106303, 2019.

\bibitem{Wang2019Blind}
H.~C. Wang, ``Blind source extraction of rolling bearings' multi-type faults
  based on self-learned sparse atomics,'' {\em Proceedings of the Institution
  of Mechanical Engineers, Part C: Journal of Mechanical Engineering Science},
  vol.~233, no.~13, pp.~4531--4542, 2019.

\bibitem{Yang2019Rolling}
R.~Yang, H.~Li, C.~Wang, and C.~He, ``Rolling element bearing weak feature
  extraction based on improved optimal frequency band determination,'' {\em
  Proceedings of the Institution of Mechanical Engineers, Part C: Journal of
  Mechanical Engineering Science}, vol.~233, no.~2, pp.~623--634, 2019.

\bibitem{Liu2018Artificial}
R.~Liu, B.~Yang, E.~Zio, and X.~Chen, ``Artificial intelligence for fault
  diagnosis of rotating machinery: A review,'' {\em Mechanical Systems and
  Signal Processing}, vol.~108, pp.~33--47, 2018.

\bibitem{Zhao2019Deep}
R.~Zhao, R.~Yan, Z.~Chen, K.~Mao, P.~Wang, and R.~X. Gao, ``Deep learning and
  its applications to machine health monitoring,'' {\em Mechanical Systems and
  Signal Processing}, 2019.

\bibitem{Lei2020Applications}
Y.~Lei, B.~Yang, X.~Jiang, F.~Jia, N.~Li, and A.~K. Nandi, ``Applications of
  machine learning to machine fault diagnosis: A review and roadmap,'' {\em
  Mechanical Systems and Signal Processing}, vol.~138, p.~106587, 2020.

\bibitem{Cortes1995Support-vector}
C.~Cortes and V.~Vapnik, ``Support-vector networks,'' {\em Machine Learning},
  vol.~20, no.~3, pp.~273--297, 1995.

\bibitem{Widodo2007Support}
A.~Widodo and B.-S. Yang, ``Support vector machine in machine condition
  monitoring and fault diagnosis,'' {\em Mechanical Systems and Signal
  Processing}, vol.~21, pp.~2560--2574, 8 2007.

\bibitem{Cover1967Nearest}
T.~Cover and P.~Hart, ``Nearest neighbor pattern classification,'' {\em IEEE
  Transactions on Information Theory}, vol.~13, no.~1, pp.~21--27, 1967.

\bibitem{Xian2009intelligent}
G.~M. Xian and B.~Q. Zeng, ``An intelligent fault diagnosis method based on
  wavelet packer analysis and hybrid support vector machines,'' {\em Expert
  Systems with Applications}, vol.~36, no.~10, pp.~12131--12136, 2009.

\bibitem{Yang2007fault}
Y.~Yang, D.~Yu, and J.~Cheng, ``A fault diagnosis approach for roller bearing
  based on imf envelope spectrum and svm,'' {\em Measurement: Journal of the
  International Measurement Confederation}, vol.~40, no.~9-10, pp.~943--950,
  2007.

\bibitem{Hao2011Application}
R.~Hao, Z.~Peng, Z.~Feng, and F.~Chu, ``Application of support vector machine
  based on pattern spectrum entropy in fault diagnostics of rolling element
  bearings,'' {\em Measurement Science and Technology}, vol.~22, no.~4, 2011.

\bibitem{Liu2013Multi-fault}
Z.~Liu, H.~Cao, X.~Chen, Z.~He, and Z.~Shen, ``Multi-fault classification based
  on wavelet svm with pso algorithm to analyze vibration signals from rolling
  element bearings,'' {\em Neurocomputing}, vol.~99, pp.~399--410, 1 2013.

\bibitem{Guo2009Rolling}
L.~Guo, J.~Chen, and X.~Li, ``Rolling bearing fault classification based on
  envelope spectrum and support vector machine,'' {\em JVC/Journal of Vibration
  and Control}, vol.~15, pp.~1349--1363, 9 2009.

\bibitem{Sharma2016Feature}
A.~Sharma, M.~Amarnath, and P.~K. Kankar, ``Feature extraction and fault
  severity classification in ball bearings,'' {\em JVC/Journal of Vibration and
  Control}, vol.~22, pp.~176--192, 1 2016.

\bibitem{Xi2000Bearing}
F.~Xi, Q.~Sun, and G.~Krishnappa, ``Bearing diagnostics based on pattern
  recognition of statistical parameters,'' {\em JVC/Journal of Vibration and
  Control}, vol.~6, pp.~375--392, 8 2000.

\bibitem{Li2020Raw}
H.~Li, Q.~Zhang, X.~Qin, and S.~Yuantao, ``Raw vibration signal pattern
  recognition with automatic hyper-parameter-optimized convolutional neural
  network for bearing fault diagnosis,'' {\em Proceedings of the Institution of
  Mechanical Engineers, Part C: Journal of Mechanical Engineering Science},
  vol.~234, no.~1, pp.~343--360, 2020.

\bibitem{Lv2021Fault}
D.~Lv, H.~Wang, and C.~Che, ``Fault diagnosis of rolling bearing based on
  multimodal data fusion and deep belief network,'' {\em Proceedings of the
  Institution of Mechanical Engineers, Part C: Journal of Mechanical
  Engineering Science}, vol.~235, no.~22, pp.~6577--6585, 2021.

\bibitem{0CWRU}
``Cwru bearing data center.''
\newblock [Online accessed 2020-08-03]
  \url{https://engineering.case.edu/bearingdatacenter}.

\bibitem{Daga2019Politecnico}
A.~P. Daga, A.~Fasana, S.~Marchesiello, and L.~Garibaldi, ``The politecnico di
  torino rolling bearing test rig: Description and analysis of open access
  data,'' {\em Mechanical Systems and Signal Processing}, vol.~120,
  pp.~252--273, 2019.

\bibitem{Nectoux2012PRONOSTIA}
P.~Nectoux, R.~Gouriveau, K.~Medjaher, E.~Ramasso, B.~Chebel-morello,
  N.~Zerhouni, and C.~Varnier, ``Pronostia : An experimental platform for
  bearings accelerated degradation tests,'' (Denver, CO, USA), 2012.

\bibitem{Qiu2006Wavelet}
H.~Qiu, J.~Lee, J.~Lin, and G.~Yu, ``Wavelet filter-based weak signature
  detection method and its application on rolling element bearing
  prognostics,'' {\em Journal of Sound and Vibration}, vol.~289, no.~4-5,
  pp.~1066--1090, 2006.

\bibitem{Smith2015Rolling}
W.~A. Smith and R.~B. Randall, ``Rolling element bearing diagnostics using the
  case western reserve university data: A benchmark study,'' {\em Mechanical
  Systems and Signal Processing}, vol.~64-65, pp.~100--131, 2015.

\bibitem{Krizhevsky2017Imagenet}
B.~A. Krizhevsky, I.~Sutskever, and G.~E. Hinton, ``Imagenet classification
  with deep convolutional neural networks,'' {\em Communications of the ACM},
  vol.~60, no.~6, pp.~84--90, 2017.

\bibitem{Fukushima1982Neocognitron}
K.~Fukushima and S.~Miyake, ``Neocognitron: A self-organizing neural network
  model for a mechanism of visual pattern recognition,'' in {\em Competition
  and cooperation in neural nets}, pp.~267--285, Springer Berlin Heidelberg,
  1982.

\bibitem{Wu2008Top}
X.~Wu, V.~Kumar, Q.~J. Ross, J.~Ghosh, Q.~Yang, H.~Motoda, G.~J. McLachlan,
  A.~Ng, B.~Liu, P.~S. Yu, Z.~H. Zhou, M.~Steinbach, D.~J. Hand, and
  D.~Steinberg, ``Top 10 algorithms in data mining,'' {\em Knowledge and
  Information Systems}, vol.~14, no.~1, pp.~1--37, 2008.

\bibitem{Alabsi2021Bearing}
M.~Alabsi, Y.~Liao, and A.~A. Nabulsi, ``Bearing fault diagnosis using deep
  learning techniques coupled with handcrafted feature extraction: A
  comparative study,'' {\em JVC/Journal of Vibration and Control}, vol.~27,
  pp.~404--414, 2 2021.

\bibitem{Guo2017Deep}
X.~Guo, C.~Shen, and L.~Chen, ``Deep fault recognizer: An integrated model to
  denoise and extract features for fault diagnosis in rotating machinery,''
  {\em Applied Sciences (Switzerland)}, vol.~7, no.~1, 2017.

\bibitem{Liu2016Rolling}
H.~Liu, L.~Li, and J.~Ma, ``Rolling bearing fault diagnosis based on stft-deep
  learning and sound signals,'' {\em Shock and Vibration}, vol.~2016, 2016.

\bibitem{Guo2018novel}
S.~Guo, T.~Yang, W.~Gao, and C.~Zhang, ``A novel fault diagnosis method for
  rotating machinery based on a convolutional neural network,'' {\em Sensors
  (Switzerland)}, vol.~18, no.~5, 2018.

\bibitem{Islam2019Automated}
M.~M. Islam and J.~M. Kim, ``Automated bearing fault diagnosis scheme using 2d
  representation of wavelet packet transform and deep convolutional neural
  network,'' {\em Computers in Industry}, vol.~106, pp.~142--153, 2019.

\bibitem{Sun2017intelligent}
W.~Sun, B.~Yao, N.~Zeng, B.~Chen, Y.~He, X.~Cao, and W.~He, ``An intelligent
  gear fault diagnosis methodology using a complex wavelet enhanced
  convolutional neural network,'' {\em Materials}, vol.~10, no.~7, 2017.

\bibitem{Wen2018New}
L.~Wen, X.~Li, L.~Gao, and Y.~Zhang, ``A new convolutional neural network-based
  data-driven fault diagnosis method,'' {\em IEEE Transactions on Industrial
  Electronics}, vol.~65, no.~7, pp.~5990--5998, 2018.

\bibitem{Janssens2018Deep}
O.~Janssens, V.~D.~R. Walle, M.~Loccufier, and V.~S. Hoecke, ``Deep learning
  for infrared thermal image based machine health monitoring,'' {\em IEEE/ASME
  Transactions on Mechatronics}, vol.~23, no.~1, pp.~151--159, 2018.

\bibitem{Yuan2018novel}
Z.~Yuan, L.~Zhang, and L.~Duan, ``A novel fusion diagnosis method for rotor
  system fault based on deep learning and multi-sourced heterogeneous
  monitoring data,'' {\em Measurement Science and Technology}, vol.~29, no.~11,
  2018.

\bibitem{Zhou2018Health}
P.~Zhou, G.~Zhou, Z.~Zhu, C.~Tang, Z.~He, W.~Li, and F.~Jiang, ``Health
  monitoring for balancing tail ropes of a hoisting system using a
  convolutional neural network,'' {\em Applied Sciences (Switzerland)}, vol.~8,
  no.~8, 2018.

\bibitem{Yoo2018novel}
Y.~Yoo and J.~G. Baek, ``A novel image feature for the remaining useful
  lifetime prediction of bearings based on continuous wavelet transform and
  convolutional neural network,'' {\em Applied Sciences (Switzerland)}, vol.~8,
  no.~7, 2018.

\bibitem{Shao2019Highly}
S.~Shao, S.~McAleer, R.~Yan, and P.~Baldi, ``Highly accurate machine fault
  diagnosis using deep transfer learning,'' {\em IEEE TRANSACTIONS ON
  INDUSTRIAL INFORMATICS}, vol.~15, no.~4, pp.~2446--2455, 2019.

\bibitem{Brusa2021Deep}
E.~Brusa, C.~Delprete, and L.~G. Di~Maggio, ``Deep transfer learning for
  machine diagnosis : From sound and music recognition to bearing fault
  detection,'' {\em Applied Sciences (Switzerland)}, vol.~11, no.~24, 2021.

\bibitem{Cao2018Preprocessing-Free}
P.~Cao, S.~Zhang, and J.~Tang, ``Preprocessing-free gear fault diagnosis using
  small datasets with deep convolutional neural network-based transfer
  learning,'' {\em IEEE Access}, vol.~6, pp.~26241--26253, 2018.

\bibitem{Xiao2019Transfer}
D.~Xiao, Y.~Huang, C.~Qin, Z.~Liu, Y.~Li, and C.~Liu, ``Transfer learning with
  convolutional neural networks for small sample size problem in machinery
  fault diagnosis,'' {\em Proceedings of the Institution of Mechanical
  Engineers, Part C: Journal of Mechanical Engineering Science}, vol.~233,
  no.~14, pp.~5131--5143, 2019.

\bibitem{Zhang2017Transfer}
R.~Zhang, H.~Tao, L.~Wu, and Y.~Guan, ``Transfer learning with neural networks
  for bearing fault diagnosis in changing working conditions,'' {\em IEEE
  Access}, vol.~5, pp.~14347--14357, 2017.

\bibitem{Zhang2021intelligent}
J.~Zhang, Q.~Zhang, X.~Qin, and Y.~Sun, ``An intelligent fault diagnosis method
  based on domain adaptation for rolling bearings under variable load
  conditions,'' {\em Proceedings of the Institution of Mechanical Engineers,
  Part C: Journal of Mechanical Engineering Science}, vol.~235, no.~24,
  pp.~8025--8038, 2021.

\bibitem{Han2021}
T.~Han, Y.~F. Li, and M.~Qian, ``{A Hybrid Generalization Network for
  Intelligent Fault Diagnosis of Rotating Machinery under Unseen Working
  Conditions},'' {\em IEEE Transactions on Instrumentation and Measurement},
  vol.~70, 2021.

\bibitem{Han2019}
T.~Han, C.~Liu, L.~Wu, S.~Sarkar, and D.~Jiang, ``{An adaptive spatiotemporal
  feature learning approach for fault diagnosis in complex systems},'' {\em
  Mechanical Systems and Signal Processing}, vol.~117, pp.~170--187, 2019.

\bibitem{Zhiyi2020}
H.~Zhiyi, S.~Haidong, J.~Lin, C.~Junsheng, and Y.~Yu, ``{Transfer fault
  diagnosis of bearing installed in different machines using enhanced deep
  auto-encoder},'' {\em Measurement: Journal of the International Measurement
  Confederation}, vol.~152, p.~107393, 2020.

\bibitem{Klein2014Bearing}
R.~Klein, E.~Masad, E.~Rudyk, and I.~Winkler, ``Bearing diagnostics using image
  processing methods,'' {\em Mechanical Systems and Signal Processing},
  vol.~45, no.~1, pp.~105--113, 2014.

\bibitem{Shao2017Rolling}
H.~Shao, H.~Jiang, F.~Wang, and Y.~Wang, ``Rolling bearing fault diagnosis
  using adaptive deep belief network with dual-tree complex wavelet packet,''
  {\em ISA Transactions}, vol.~69, pp.~187--201, 7 2017.

\bibitem{Brunton2019Data}
S.~L. Brunton and J.~N. Kutz, {\em Data Driven Science and Engineering: Machine
  Learning, Dynamical Systems, and Control}.
\newblock Cambridge University Press, 2019.

\bibitem{Belhumeur1997Eigenfaces}
P.~N. Belhumeur, J.~P. Hespanha, and D.~J. Kriegman, ``Eigenfaces vs.
  fisherfaces: Recognition using class specific linear projection,'' {\em IEEE
  Transactions on Pattern Analysis and Machine Intelligence}, vol.~19, no.~7,
  pp.~711--720, 1997.

\bibitem{Kirby1990Application}
M.~Kirby and L.~Sirovich, ``Application of the karhunen-loéve procedure for
  the characterization of human faces,'' {\em IEEE Transactions on Pattern
  Analysis and Machine Intelligence}, vol.~12, no.~1, pp.~103--108, 1990.

\bibitem{Sirovich1987Low-dimensional}
L.~Sirovich and M.~Kirby, ``Low-dimensional procedure for the characterization
  of human faces,'' {\em J. Opt. Soc. Am. A}, vol.~4, pp.~519--524, 3 1987.

\bibitem{Turk1991Eigenfaces}
M.~Turk and A.~Pentland, ``Eigenfaces for recognition,'' {\em Journal of
  cognitive neuroscience}, vol.~3, no.~1, pp.~71--86, 1991.

\bibitem{Tian2015Big}
W.~Tian and Y.~Zhao, ``Big data technologies and cloud computing,'' in {\em
  Optimized Cloud Resource Management and Scheduling}, pp.~17--49, Elsevier, 1
  2015.

\bibitem{Wang2013enhanced}
D.~Wang, P.~W. Tse, and K.-L. Tsui, ``An enhanced kurtogram method for fault
  diagnosis of rolling element bearings,'' {\em Mechanical Systems and Signal
  Processing}, vol.~35, pp.~176--199, 2 2013.

\bibitem{Wang2011adaptive}
Y.~Wang and M.~Liang, ``An adaptive sk technique and its application for fault
  detection of rolling element bearings,'' {\em Mechanical Systems and Signal
  Processing}, vol.~25, no.~5, pp.~1750--1764, 2011.

\bibitem{Halko2011Finding}
N.~Halko, P.~G. Martinsson, and J.~A. Tropp, ``Finding structure with
  randomness: Probabilistic algorithms for constructing approximate matrix
  decompositions,'' {\em SIAM Review}, vol.~53, no.~2, pp.~217--288, 2011.

\bibitem{Martinsson2011randomized}
P.~G. Martinsson, V.~Rokhlin, and M.~Tygert, ``A randomized algorithm for the
  decomposition of matrices,'' {\em Applied and Computational Harmonic
  Analysis}, vol.~30, no.~1, pp.~47--68, 2011.

\bibitem{Sarlos2006Improved}
T.~Sarlós, ``Improved approximation algorithms for large matrices via random
  projections,'' {\em Proceedings - Annual IEEE Symposium on Foundations of
  Computer Science, FOCS}, pp.~143--152, 2006.

\bibitem{Erichson2019Randomized}
N.~B. Erichson, S.~Voronin, S.~L. Brunton, and J.~N. Kutz, ``Randomized matrix
  decompositions using r,'' {\em Journal of Statistical Software}, vol.~89,
  2019.

\bibitem{Pearson1901On}
K.~Pearson, ``On lines and planes of closest fit to systems of point in
  space,'' {\em Philosophical Magazine}, vol.~2, no.~11, pp.~559--572, 1901.

\bibitem{Zhang2020Machine-learning-based}
G.~Zhang, L.~Tang, Z.~Liu, L.~Zhou, Y.~Liu, and Z.~Jiang,
  ``Machine-learning-based damage identification methods with features derived
  from moving principal component analysis,'' {\em Mechanics of Advanced
  Materials and Structures}, vol.~27, no.~21, pp.~1789--1802, 2020.

\bibitem{Escalera2009}
S.~Escalera, O.~Pujol, and P.~Radeva, ``{Separability of ternary codes for
  sparse designs of error-correcting output codes},'' {\em Pattern Recognition
  Letters}, vol.~30, no.~3, pp.~285--297, 2009.

\bibitem{Li2013}
W.~Li, S.~Zhang, and G.~He, ``{Semisupervised Distance-Preserving
  Self-Organizing Map for Machine-Defect Detection and Classification},'' {\em
  IEEE Transactions on Instrumentation and Measurement}, vol.~62, no.~5,
  pp.~869--879, 2013.

\bibitem{Zhang2015}
X.~Zhang, Y.~Liang, J.~Zhou, and Y.~Zang, ``{A novel bearing fault diagnosis
  model integrated permutation entropy, ensemble empirical mode decomposition
  and optimized SVM},'' {\em Measurement: Journal of the International
  Measurement Confederation}, vol.~69, pp.~164--179, 2015.

\bibitem{Lei2016}
Y.~Lei, F.~Jia, J.~Lin, S.~Xing, and S.~X. Ding, ``{An Intelligent Fault
  Diagnosis Method Using Unsupervised Feature Learning Towards Mechanical Big
  Data},'' {\em IEEE Transactions on Industrial Electronics}, vol.~63, no.~5,
  pp.~3137--3147, 2016.

\bibitem{Ding2017}
X.~Ding and Q.~He, ``{Energy-Fluctuated Multiscale Feature Learning with Deep
  ConvNet for Intelligent Spindle Bearing Fault Diagnosis},'' {\em IEEE
  Transactions on Instrumentation and Measurement}, vol.~66, no.~8,
  pp.~1926--1935, 2017.

\bibitem{Zhao2018}
R.~Zhao, D.~Wang, and R.~Yan, ``{Machine Health Monitoring Using Local
  Feature-Based Gated Recurrent Unit Networks},'' {\em IEEE Transactions on
  Industrial Electronics}, vol.~65, no.~2, pp.~1539--1548, 2018.

\end{thebibliography}


\end{document}